\documentclass[12pt,preprint]{aastex}

\shorttitle{Intrinsic Extreme Ultraviolet Fluxes}
\shortauthors{Linsky et al.}
\slugcomment{Draft \today}

\begin{document}

\newcommand{\php}[0]{\phantom{--}}
\newcommand{\kms}[0]{km~s$^{-1}$}

\title{THE INTRINSIC EXTREME ULTRAVIOLET FLUXES\\ 
OF F5 V TO M5 V STARS}

\author{Jeffrey L. Linsky\altaffilmark{1}}
\affil{JILA, University of Colorado and NIST, 440UCB Boulder, CO 80309-0440, 
USA}
\email{jlinsky@jilau1.colorado.edu}

\author{Juan Fontenla}
\affil{NorthWest Research Associates Inc., 3380 Mitchell Ln, Boulder CO 80301, 
USA}
\email{ jfontenla@nwra.com}

\and

\author{Kevin France\altaffilmark{1,2}}
\affil{CASA, University of Colorado, 593UCB Boulder, CO 80309-0593, USA}
\email{Kevin.France@colorado.edu}



\altaffiltext{1}{User of the MAST Data Archive at the Space Telescope 
Science Institute. STScI is operated by the Association of Universities for 
Research in Astronomy, Inc., under NASA contract NAS 5-26555.}

\altaffiltext{2}{NASA Nancy Grace Roman Fellow.}

\begin{abstract}

\noindent Extreme ultraviolet (EUV) radiations (10--117~nm) from host stars 
play important roles
in the ionization, heating, and mass loss from exoplanet atmospheres. Together 
with the host star's Ly$\alpha$ and far-UV (117--170~nm) radiation, 
EUV radiation 
photodissociates important molecules, thereby changing the chemistry
in exoplanet atmospheres. Since stellar EUV fluxes cannot now be measured 
and interstellar neutral hydrogen completely 
obscures stellar radiation between 40 and 91.2~nm, even for the nearest stars, 
we must estimate the unobservable EUV flux by 
indirect methods. New non-LTE semiempirical models of the solar chromosphere 
and corona and solar irradiance measurements show that the ratio of EUV flux 
in a variety of wavelength bands to the Ly$\alpha$ flux varies slowly 
with the Ly$\alpha$ flux and thus with the magnetic heating rate. 
This suggests and we confirm that solar EUV/Ly$\alpha$ flux ratios 
based on the models and observations are similar to 
the available 10--40~nm flux ratios observed with 
the {\em EUVE} satellite and the 91.2--117~nm flux
observed with the {\em FUSE} satellite for F5~V--M5~V stars. We provide 
formulae for predicting EUV flux ratios based on the
{\em EUVE} and {\em FUSE} stellar data and on the solar models,
which are essential input for modelling the atmospheres of exoplanets.

\end{abstract}


\section{INTRODUCTION}

The discovery of many extrasolar planets (exoplanets) by 
radial velocity, transit, and imaging techniques has stimulated observational 
and theoretical studies to characterize their atmospheric chemistry and 
physical properties and to investigate whether these exoplanets could sustain
life forms \citep[e.g.,][]{Kasting2003,Seager2010}. As density decreases
with height in exoplanet atmospheres, photolysis (photodissociation of 
molecules and photoionization of atoms) will eventually dominate 
over thermal equilibrium. This typically occurs where the atmospheric
pressure is less than 1 mbar. 
Recently photochemical models have been computed for 
terrestrial planets and super-Earths 
\citep[e.g.,][]{Segura2010,Kaltenegger2011,Hu2012},
hot-Neptunes \citep{Line2011}, and
hot-Jupiters \citep{Kopparapu2012,Moses2013,Line2010}.
Far Ultraviolet (FUV) radiation at wavelengths below 170~nm and, in 
particular, the very bright Ly$\alpha$ emission line (121.6~nm), 
control the 
photodissociation of such important molecules as H$_2$O, CH$_4$, and CO$_2$,
which can increase the mixing ratio of oxygen \citep{Tian2013}. 
Ozone (O$_3$) has been called a potential biosignature in
super-Earth atmospheres \citep{Segura2005, Segura2010, Grenfell2013}, but
it is important to assess the extent to which photolysis of O$_2$ and 
subsequent chemical reactions rather than biological processes can control 
its abundance. Future photochemical models based on realistic host star 
UV emission including intrinsic Ly$\alpha$ fluxes are needed to address
questions of the reliability of biosignatures and atmospheric chemical 
abundances. Recent models, such as those 
cited above, show that the C/O ratio, quenching reactions, 
thermal structure, and diffusion also play important roles in determining 
mixing ratios for important molecules in exoplanet atmospheres, but the
short wavelength radiation of the host star is critically important.

Atmospheric chemistry models require as input the FUV 
(117--170~nm) radiation from the host star. Spectra obtained with the
Cosmic Origins Spectrograph (COS) and Space Telescope Imaging Spectrograph 
(STIS) instruments on {\em Hubble Space Telescope (HST)} are providing 
these data 
\citep[e.g.,][]{Ayres2010} including M dwarf stars \citep{France2013},
which many authors believe are the most favorable candidate host stars with
nearby Earth-like exoplanets \citep{Scalo2007,Tarter2007}. 
The {\em Galaxy Evolution Explorer (GALEX)} instrument is also providing 
broadband FUV (not including the 
Ly$\alpha$ line) and NUV fluxes
of exoplanet host stars \citep{Shkolnik2013}.
While the Ly$\alpha$
line is the most important FUV emission feature for solar-type stars and 
is as bright as the entire 120--320~nm spectrum of M dwarfs \citep{France2013},
the entire core of this line is absorbed by interstellar hydrogen. The
intrinsic flux in the Ly$\alpha$ line can be reconstructed from 
high-resolution spectra \citep{Wood2005,France2013} or predicted from
correlations with other emission lines \citep{Linsky2013}.

At wavelengths of 10--91.2~nm, extreme-ultraviolet (EUV) radiation 
from the host star photoionizes hydogen creating an ionosphere 
\citep{Koskinen2010} and heats the outer layers of these atmospheres, thereby 
inflating the atmosphere and driving mass loss. 
\citet{Murray2009} computed models that
describe how photoionization heating of hot-Jupiter atmospheres by EUV 
radiation drives transonic hydrodynamic outflows (also called hydrodynamic 
blow-off). These outflows are analogous to the Parker-type 
solar wind \citep{Parker1958}, except that the heating is from above 
rather than below. For a hot-Jupiter exoplanet like HD~209458b located at 
0.05~AU from its solar-type host star, the outflowing plasma is heated almost 
entirely by the kinetic energy of protons (and their subsequent collisions) 
after hydrogen atoms are ionized by host star EUV photons with a small 
contribution of X-ray photons at $\lambda<10$~nm \citep{Koskinen2010}. 
Hot-Jupiters and Neptune-like exoplanets with hydrogen-rich atmospheres and 
weak magnetic moments can also lose mass when photoionized and charge-exchanged
atomic and molecular hydrogen are picked-up by the stellar wind and 
coronal mass ejection 
plasma outside of the magnetopause standoff distance where 
the stellar wind pressure exceeds the planet's magnetic pressure
\citep[e.g.,][]{Lammer2003,Griessmeier2004,Khodachenko2007,Vidotto2011}.

The expanding atmospheres of exoplanets in orbit around older solar-like stars 
are heated by the incident EUV radiation and cooled by
expansion (PdV work). These winds are described as {\em energy limited}.
\citet{Murray2009} showed that the mass-loss rates of such winds are 
proportional to the incident EUV flux, $f_{\rm EUV}^{0.9}$. 
When the host star has a far larger
EUV flux, for example a T Tauri star, X-ray heating also becomes important 
and radiative recombination rather than 
expansion cools the denser wind. The mass-loss rate for such winds
is proportional to $f_{\rm EUV}^{0.6}$ and \citet{Murray2009} call such winds  
{\em radiation/recombination-limited}. They argue that the proper way to 
describe mass loss from the hydrogen-rich atmospheres of exoplanets 
located close to their host stars is by transonic hydrodynamic winds 
heated by EUV photons rather than by radiation pressure 
\citep{Vidal-Madjar2003}. \citet{Yelle2004}, \citet{Tian2005}, 
\citet{Garcia2007}, and  others have also computed transonic hydrodynamic
outflows for hot-Jupiters, and \citet{Lammer2013} have computed such models
for super-Earths with hydrogen-rich upper atmospheres. Super-Earths have much 
smaller mass loss rates than hot-Jupiters. Roche lobe
overflow can enhance the mass loss rate \citep{Erkaev2007}, but the ram and 
magnetic pressure of a strong stellar wind 
on the day side of the exoplanet can suppress a transonic outflow, 
producing instead a subsonic outflow often called a Jeans-type stellar breeze.
The supersonic orbital speed of a hot-Jupiter moving through a stellar wind 
can produce a nonspherical shock front ahead of the exoplanet's motion with
properties that depend on the stellar wind, magnetic field, and exoplanet's
mass loss rate \citep[e.g.,][]{Bisikalo2013}.
For all of these cases, the host star's unobservable EUV flux and the 
model-dependent fraction of this flux that is converted to heat 
\citep[e.g.,][]{Lammer2013} are essential input 
parameters for computing realistic models of exoplanet atmospheres.

\citet{Tian2008} calculated the response of the Earth's 
oxygen and nitrogen atmosphere to changes in the EUV flux from its host star, 
the present day and the young Sun. With a 1-D hydrodynamic model coupled to 
a code that describes EUV photoionization and heating by secondary 
electrons, they find that illumination by the present day Sun predicts that 
the upper thermosphere is in hydrostatic equilibrium, but an increase in the 
EUV flux by only a factor of 4.6 is sufficient to produce 
a hydrodynamic outflow that becomes the dominant cooling mechanism. 
A factor of 10 increase in the EUV flux predicted for the young Sun 
produces a transonic outflow.

Estimating the EUV emission of host stars is, therefore, an essential but 
difficult problem to solve because interstellar hydrogen absorbs
essentially all of the spectrum between 40 and 91.2~nm, even for the nearest 
stars, and there are spectra in the 10--40~nm range for only a 
few stars observed with the {\em Extreme Ultraviolet Explorer (EUVE)} satellite
\citep{Craig1997,Sanz-Forcada2003}.

Despite these problems, several authors have developed methods for estimating
the EUV flux for solar-type stars.
\citet{Ayres1997} estimated EUV fluxes and photodissociation rates from the 
observed FUV and X-ray fluxes of solar-type stars and discussed the effect
of enhanced EUV emission from the young Sun on the early Martian atmosphere.
Following a similar approach, \citet{Ribas2005,Ribas2010} estimated EUV fluxes 
for solar-type stars from the FUV and X-ray emission of the Sun and six stars 
with spectral types G0~V to G5~V and a range of ages and thus activity. 
This work is appropriate for solar-type 
stars, but its applicability to other spectral type stars is not 
discussed in their papers. Recently, \citet{Claire2012} developed a technique 
for estimating the EUV to IR flux of the Sun as a function of age 
(0.6--6.7 Gyr) by 
computing relative flux multipliers for different wavelength intervals 
using observations of the G-type stars $\kappa$~Cet and EK~Dra to test the 
multipliers at earlier ages when the Sun was more active. However, 
\citet{Claire2012} do not 
extend their approach to estimating fluxes for stars much different in
spectral type from the Sun.

\citet{Sanz-Forcada2011} computed synthetic EUV 
spectra of many F--M stars from emission measure distribution analyses of 
their X-ray spectra. The EUV fluxes computed from their
synthetic spectra may not accurately include emission in the hydrogen 
Lyman continuum,
important for the 70--91.2~nm region, the He~I and He~II continua, 
and may exclude some of the 
emission lines seen in solar spectra. We will compare our results with those
of \citet{Sanz-Forcada2011} later in this paper.

In this paper our objective is to develop a different kind of 
technique for estimating 
the EUV emission of host stars with spectral types F--M that is relatively 
insensitive to stellar activity and variability. Our technique is similar
to that used by \citet{Linsky2013} in that we estimate the ratios of EUV 
fluxes in different wavelength bands to the emission in a representative 
emission line, in this case Ly$\alpha$. Since both the EUV and 
Ly$\alpha$ fluxes increase and decrease together (but not necessarily 
at the same rate) with the magnetic heating 
rate that depends on stellar rotation, age, and magnetic field properties, 
EUV/Ly$\alpha$ flux ratios should change rather slowly with the 
Ly$\alpha$ flux. In support of this hypothesis, \citet{Claire2012} show
that the number of Ly$\alpha$ photons equals the total number of photons
emitted below 170~nm by the Sun at all ages from the zero age main sequence to 
the present, while the ratio of Ly$\alpha$ flux to the total solar flux 
below 170~nm increases smoothly from 20 to 36.5\% over this time interval. 
We also find that EUV/Ly$\alpha$ flux ratios vary slowly with activity 
and that the flux ratios in solar 
data and recent solar irradiance models are representative of stars
with a wide range of spectral type and activity. Observations with the
{\em EUVE} satellite in the 10--40~nm wavelength 
range and the {\em Far Ultraviolet  Spectroscopic Explorer (FUSE)} satellite 
in the 91.2--117~nm range provide empirical tests of our method.
When the reconstructed Ly$\alpha$ flux is not available for a given star, 
it may be estimated using the techniques described by \citet{Linsky2013}.

\section{SOLAR AND STELLAR EUV SPECTRA}

A natural division between the EUV and the FUV occurs at 117~nm where the 
reflectivity of Al+MgF$_2$-coated optics of {\em HST} and other spectrographs
rapidly falls to low values. Efficient observations at shorter wavelengths 
require different optical coatings or grazing incidence optics.
We therefore consider the EUV to extend from 10--117~nm and the FUV to
extend from 117--170~nm. The C~III 117.7~nm multiplet is in the overlap region 
observed by both {\em FUSE} and {\em HST}.

\subsection{\it Solar Irradiance Reference Data}

The Sun is the only star for which the EUV spectrum can be observed in its 
entirety without attenuation by the interstellar medium (ISM).
Since interstellar absorption prevents detection of much of the EUV
spectrum for even the nearest stars, in particular the 40--91.2~nm portion 
of the spectrum, 
we begin this study with the analysis of the solar irradiance spectrum, the 
flux of the Sun observed as a star. The solar irradiance spectrum has been the
subject of intense study stimulated, in large part, by the need to determine 
its variability on short- and long-period time scales that could influence the 
chemical composition of the Earth's atmosphere and possibly drive 
terrestrial climate change. \citet{Woods2002} have reviewed the earlier
studies of solar UV and EUV irradiance variability and pointed out the 
developing instrumental techniques for increasing the photometric 
accuracy of the data.
 
Figure 1 shows the 
composite irradiance spectrum at solar minimum (March--April 2008, Carrington
Rotation 2068) when the solar 10.7~cm flux was at a very low value of 
$69\times 10^{-22}$ W m$^{-2}$ Hz$^{-1}$. At this time, an observing campaign 
with four instruments in space \citep{Woods2009} measured the Solar 
Irradiance Reference Spectrum (SIRS) between 0.1 and 2400~nm, which 
was kindly provided by Martin Snow. The 0.1--6.0~nm spectrum was observed with 
the XUV (soft X-ray) Photometer System (XPS) of the Solar EUV Experiment (SEE) 
\citep{Woods2005} on the {\em Thermosphere, Ionosphere, Mesosphere, 
Energetics, and Dynamics (TIMED)} spacecraft. The rocket prototype 
EUV Variability
Experiment (EVE) \citep{Chamberlin2009} monitored the 
6.0--105~nm spectral interval, and the EUV Grating Spectrograph (EGS)
on SEE obtained the 105--116~nm spectral interval. The 116--310~nm
spectrum was measured by the Solar Radiation and Climate Experiment (SORCE)
spectrometer on the {\em Solar Stellar Irradiance Comparison Experiment 
(SOLTICE)} satellite \citep{McClintock2005}.

Tom Woods has kindly provided a second set of solar irradiance data for
times of solar minimum and maximum. These data were obtained with the
SEE instrument on {\em TIMED} using the version 11 calibration.
The solar minimum data are for day 105 in 2008, and the solar maximum 
data are for day 76 in 2002. We refer to this data set as the SEE data.
Table~1 lists the SIRS and SEE fluxes in different wavelength bands.

Despite very careful contamination control and calibration before launch 
\citep[e.g.,][]{Chamberlin2009,Hock2012},
EUV spectrometers pointed at the Sun for long periods 
of time typically show sensitivity degradation due to contaminated optics
and detector aging \citep{BenMoussa2013}. The 6.0--105~nm data in the SIRS
data set should show minimal degradation as the spectrometer flew on
a rocket and was calibrated before and after flight. The SEE 
data set, however, could include larger degradation, which is difficult to 
calibrate, as the solar minimum and maximum data
were obtained six years apart with the same instrument in orbit.

\subsection{\it Stellar 91.2--117.0~nm Spectra}

With its LiF and SiC overcoated optics, the {\em Far 
Ultraviolet Spectrograph Explorer (FUSE)} satelite was able to observe 
nearby stars at wavelengths between the
Lyman continuum bound-free edge at 91.2~nm and 117.0~nm. 
For a description of the {\em FUSE} satellite and observing program see
\citet{Moos2000} and \citet{Sahnow2000}. \citet{Redfield2002}
described {\em FUSE} spectra of seven A7~V to M0~V stars, and 
\citet{Dupree2005} described {\em FUSE} spectra of eight F--M giants. 
The 91.2--117.0~nm spectrum
is dominated by emission in Ly$\beta$ and higher Lyman lines 
(hereafter called the Lyman series) and emission lines of C~II 103.6~nm, 
C~III 97.7~nm, and O~VI 103.2 and 103.8~nm. 
The Lyman lines are formed in the chromosphere
at $\log T\approx 3.8$, C~II lines near the base of the transition region 
($\log T\approx 4.3$), and the C~III and O~VI lines in the transition region at
$\log T\approx 4.8$ and 5.5, respectively. Much of the flux in the Lyman 
lines is absorbed or scattered by interstellar H~I, 
and the lines are contaminated by terrestrial airglow emission
\citep{Feldman2001} that could not be removed accurately from the {\em FUSE} 
data. The weak continuum of F--G stars could not be measured by {\em FUSE}. 

The quiet Sun spectrum described above likely provides a reliable census of
the emission lines that dominate this portion of the spectrum for F--G stars. 
The brightest emission line in the 91.2--117.0~nm spectrum of the quiet Sun 
is the C~III 97.70~nm line with flux  of 
0.101 erg cm$^{-2}$ s$^{-1}$ at 1~AU. The next brightest line is Ly$\beta$ 
with 0.0655 erg cm$^{-2}$ s$^{-1}$. The total flux in the 
Lyman series is only 0.114 in these units, whereas the sum of the fluxes in 
the C~II, C~III, and O~VI lines is 0.178 in the same units. Since transition
region lines are important contributors to the 91.2--117~nm flux of the quiet
Sun, it is a sensible assumption, as confirmed by {\em FUSE} spectra, 
that the same transition region lines will also be important in this wavelength
interval in F--M dwarfs stars. However, the relative strength of transition
region lines may depend on stellar spectral type and activity.
For example, the O~VI lines are fainter than the C~III line for the F and G
stars Procyon and $\alpha$~Cen~A, are comparable in brightness for the K dwarfs
$\alpha$~Cen~B and $\epsilon$~Eri, and are brighter
than the C~III line for the active M dwarf AU~Mic \citep{Redfield2002}. 
There are also two coronal emission lines in
this spectral range, Fe~XVIII 97.486~nm and Fe~XIX 111.806~nm, but these lines
are very weak compared to the transition region lines \citep{Redfield2003}.

\citet{Redfield2002} provided a list of emission line fluxes, except for the 
Lyman series lines, for seven dwarf stars. 
Five of these stars have intrinsic Ly$\alpha$ fluxes measured
by \citet{Wood2005}: Procyon (F5~IV-V), $\alpha$~Cen~A (G2~V), 
$\alpha$~Cen~B (K0~V), $\epsilon$~Eri (K2~V), and AU~Mic (M0~V). 
The sums of these emission line fluxes, except for the 
C~III 117.5~nm blend that is observable by {\em HST}, are listed in 
Table~2.

\subsection{\it Stellar 10--40 nm Spectra}

The {\em Extreme Ultraviolet Explorer (EUVE)} obtained spectra of nearby 
stars in the 7--76~nm wavelength range with 0.05--0.2~nm spectral resolution.
For a description of the {\em EUVE} science instruments, see 
\citet{Bowyer1991} and \citet{Welsh1990}. \citet{Craig1997} presented
{\em EUVE} spectra for a variety of stars including several 
single and binary dwarf stars with F--M spectral types. 
\citet{Sanz-Forcada2003} measured the emission line fluxes of many of these 
stars between 8 and 36~nm. We obtained calibrated {\em EUVE} spectra of 
15 F--M dwarf stars with usable spectra between 10 and 40~nm
from the Mikulski Archive for Space 
Telescopes (MAST) housed at the Space Telescope Science Institute (STScI).
We downloaded only nighttime data for which scattered sunlight including 
geocoronal emission in the He~II 30.4~nm line should be minimal. 
The stars selected (see Table~3)
have good S/N, intrinsic Ly$\alpha$ fluxes \citep{Wood2005,Linsky2013},
and interstellar hydrogen column densities, log[N(HI)] \citep{Wood2005}.
In a few cases, we have estimated intrinsic Ly$\alpha$ fluxes using 
correlations with other emission lines (e.g., Mg~II, Ca~II, and C~IV)
\citep{Linsky2013}. For a few stars we have also estimated interstellar 
hydrogen column densities using similar sight lines \citep{Redfield2008}.
Estimated parameters are listed in parentheses.

Table~3 summarizes the flux ratios in the 10--20, 20--30, and 30--40~nm bands 
that we obtained from the {\em EUVE} data. Listed in the Table are the 
{\em EUVE} data identifiers, {\em EUVE} observing times, spectral types, 
intrinsic Ly$\alpha$ 
fluxes, hydrogen column densities, and ratios of the {\em EUVE} flux in three 
wavelength bands to the Ly$\alpha$ flux before ($R$) and after 
correction for interstellar absorption ($R_{\rm ISM}$), using the interstellar
absorption cross section formula of \citet{Morrison1983} computed for each 
wavelength. One {\em EUVE} 
spectrum of AU~Mic (au\_mic\_\_9207141227N) is far brighter than the
other two and contains a very large flare analyzed by \citet{Monsignori1996}
and \citet{Cully1993}.
The other two observations of AU Mic show much weaker 
emission lines, and we assume represent the star's quiescent emission. 
The {\em EUVE} spectrum of EV Lac also contains a large flare 
\citep{Mullan2006}. We have averaged two 
spectra of YZ~CMi, two nonflare spectra of AU~Mic, and 6 spectra of AD~Leo 
that do not show evidence of large flares.

\section{SOLAR MODELS}

In this paper, we use the EUV fluxes computed with the new semiempirical 
solar models of \citet{Fontenla2013}, which revise the chromosphere, 
transition region, and coronal structures of the earlier models of 
\citet{Fontenla2009} and \citet{Fontenla2011}. The new models include updated 
collisional rates, ionization equilibria, and more levels and spectral lines 
from CHIANTI 7.1 \citep{Landi2013}.
These are 1-dimensional non-LTE models of temperature vs. height structures 
selected to best fit the observed emitted intensity and spectral irradiance 
from the EUV to the infrared. The calculations in the updated models 
include 51 species of 21 elements at various stages of ionization and 
H$^-$. For the higher ionization stages, the level populations are computed 
using an "effectively optically thin" approach, but optical thickness is 
considered for some lines-of-sight.

The updated set of nine models is defined by corresponding levels of 
magnetic heating 
as observed in chomospheric, transition-region, and coronal emissions 
(e.g., Ca II H and K lines, the 1600~\AA\ continuum, and {\em SOHO/EIT} and 
{\em SDO/AIA} images). These models range from minimal activity 
(feature A represents quiet-Sun inter-network regions) to maximum activity 
(feature Q represents very hot plage). The solar feature designation (letter) 
and current photosphere-chromosphere-transition region (below 200,000~K)
model index are, in order of increasing brightness, 1300 for A, 
1301 for B, 1302 for D, 1303 for F, 1304 for H, 1305 for P, and 1308 for Q.
The corresponding models for temperatures above 200,000~K including the corona
are 1310--1318. and combination of the lower and higher temperature models
are designated 13x0--13x8. 
In the earlier models, sunspot umbra and penumbra were included, but the 
updates were made only for the models listed above, which are the important 
ones for the EUV and FUV radiation. The solar spectral irradiance, 
the solar flux  at 1~AU, is obtained from radiative transfer calculations of 
the intensity at 10 positions across the solar disk. The synthesized 
quiet-Sun computed spectrum matches the 116--168~nm irradiance at solar 
minimum activity measured by the {\em SORCE/SOLSTICE} instrument 
\citep{Woods2009} and the 6--105~nm flux obtained by the {\em EVE} 
instrument \citep{Chamberlin2009}. \citet{Fontenla2013} conclude that the
higher computed continuum in the 168-200~nm range compared to observations
is due to a missing opacity source that is likely molecular. 
Computed fluxes of most important chromospheric and transition region 
emission lines are consistent with observations, although fine details of 
the lines are not perfectly matched by these very simplified models. 
Overall, the match to the observed EUV spectra is very good, as will be 
described later.

Since the \citet{Fontenla2013} models refer to the same star but with 
different levels of EUV and UV emission indicative of different levels of 
magnetic heating (often called ``activity''),
these models are very useful for studying correlations of EUV emission with
many emission lines formed in the chromosphere and transition region as a 
function of activity for solar-like stars. \citet{Linsky2012} showed that the 
115--150~nm continuum emission from Models 1001 through 1005 
\citep{Fontenla2011} corresponds 
to the observed continuum emission from low activity old solar-mass 
stars ($\alpha$~Cen~A) to high activity young solar-mass stars
(EK Dra and HII~314). 

\section{PREDICTING STELLAR FLUXES FROM CORRELATIONS WITH LY$\alpha$}

\subsection{\it The 91.2--117~nm Portion of the EUV Spectrum}

We need to add the flux in the Lyman series lines beginning with Ly$\beta$ 
to the other emission lines for the five stars measured by 
\citet{Redfield2002}. \citet{Lemaire2012} and 
previous authors that they cite noted that the Ly$\beta$/Ly$\alpha$ 
flux ratio increases with solar activity. This is likely due to the 
higher temperatures and thus higher collisional 
excitation rates in more active 
regions on the Sun. Flux in the other Lyman series lines should also increase
faster than  Ly$\alpha$ for the same reason. 
Figure~2 shows that the
Lyman series/Ly$\alpha$ flux ratio increases from the least active
area of the Sun \citep{Fontenla2013} model 13x0 to the most active area 
(model 13x8). We fit these data with a power-law,
log[f(Lyman series)/f(Ly$\alpha$)] = A + B log[f(Ly$\alpha$)], where 
A=--1.798 and B=0.351. Table~2 shows the sum of the emission line fluxes
measured by \citet{Redfield2002}, the estimated Lyman series flux using the 
above relation, the total flux in the 91.2--117.0~nm band, and the ratio of
this flux to f(Ly$\alpha$).

Figure~3 shows the ratio of the total 91.2--117.0~nm flux to the Ly$\alpha$
flux (the EUV flux ratio) for the solar models and the solar and 
stellar data. The asterix symbols and solid line connecting them in Figure~3 
are the EUV flux ratios obtained from the \citet{Fontenla2013} semiempirical 
models 1300 to 1308.
The Sun symbol is for the observed quiet Sun ratio in the SIRS
data set, and the ``m'' and ``M'' symbols refer to the solar minimum and 
maximum data for the SEE data set. We note that the solar data lie only about 
0.10 dex below the quiet and moderately active solar models 1300--1302.
 
Since we are comparing EUV flux ratios among stars with different radii, 
we plot the EUV flux ratios in this and subsequent figures, vs. the stellar 
Ly$\alpha$ flux at 1~AU multiplied by the scale factor
$(R_{Sun}/R_{star})^2$. This scale factor enables us to compare the EUV flux
ratios to the Ly$\alpha$ flux per unit area of the stellar surface, which
is a physical measure of the chromospheric heating rate.
There is no need to scale either the EUV flux or Ly$\alpha$ flux when 
computing the EUV flux ratios as both quantities are 
proportional to the stellar radius and the ratio is thus 
independent of stellar radius. 

The flux in the Lyman series lines beginning with Ly$\beta$ 
is only about 20\% of the total 91.2--117.0~nm 
flux for the quiet Sun models, but it increases to 30\% 
for the active Sun models. This suggests that similar ratios likely apply to 
stars with similar activity levels as measured by the Ly$\alpha$ 
flux. We therefore add the estimated Lyman series fluxes to the observed
91.2--117.0~nm line fluxes for the five stars observed by {\em FUSE} and 
divide the sums by the reconstructed Ly$\alpha$ fluxes \citep{Linsky2013}.
The error bars for each star are estimates that include the 
estimated Lyman series fluxes, assuming errors in both dimensions of 
$\pm 15$\%.

The least-squares fit power law to flux ratio vs. scaled Ly$\alpha$ flux 
for the five stars and the Sun is
log[f(91.2--117.0~nm)/f(Ly$\alpha$)] = C + D log[f(Ly$\alpha$)],
where C=$-1.189 \pm 0.202$ and D=$+0.110 \pm 0.152$. Since the uncertainty
in the linear coefficient is larger than its value, we instead fit the data
with a constant value  log[f(91.2--117.0~nm)/f(Ly$\alpha$)] = -1.025.
The mean dispersion of the solar and stellar data about this relation
is only 29.5\% and the rms dispersion is 35.0\%.
Since this fit (dash-dot line) in Figure~3 provides an excellent fit 
to the solar and  F5 IV-V to M0~V stellar data, we argue that it 
can be used to predict the 91.2--117.0~nm flux from
a wide range of late-type dwarf stars, provided one has measurements of the 
Ly$\alpha$ flux or another spectral line that is correlated with
the Ly$\alpha$ flux. Note that the empirical fit is an excellent match to 
the solar data and to $\alpha$~Cen~A, which is a close match to the Sun. The
solar models (solid line in Figure~3) lie only 0.1--0.2 dex above the 
empirical fit to the solar and stellar data. While
the contribution of the 91.2--117.0~nm band to the total
flux incident on an exoplanet is relatively small compared to the flux in 
Ly$\alpha$, it will be important for the photodissociation of molecules 
that have peak cross-sections in this wavelength range (e.g., CO and H$_2$).

\subsection{\it The Hydrogen Lyman Continuum 60--91.2~nm}

The solar spectrum (Figure~1) shows that the hydrogen Lyman continuum emission
extends from the 91.2~nm edge down to nearly 60~nm. The brightest emission 
lines superimposed on the Lyman continuum are lines of 
O~III (near 84 and 72~nm), O~IV (near 79~nm), Ne~IV and Ne~VIII (76--78~nm),
and a mixture of other transition region lines at shorter wavelengths. 
We now consider how to isolate the Lyman continuum component of the EUV 
spectrum and then compare its flux to observable emission features. 

Figure~4 shows the SIRS Lyman continuum flux measured at 12 wavelengths where
there is no obvious blending with weak emission lines. The solid line is a 
least-squares linear fit in this semilog plot with 
$f(\lambda) = 5.85\times 10^{-11} e^{-0.149\lambda}$ erg~cm$^{-2}$~s$^{-1}$,
where $\lambda$ is in nm. The fit is very good with an rms deviation of 3.9\%.
\citet{Parenti2005} have previously shown that an exponential provides a very 
good fit to the Lyman continuum slope in the SOHO/SUMER radiance data. The
integrated flux in the Lyman continuum between 60 and 91.2~nm is
0.307 erg~cm$^{-2}$~s$^{-1}$ at 1~AU or 5.16\% that of the Ly$\alpha$ flux
in the same data set. By comparison, the total flux in the Lyman line series
is only 0.114  erg~cm$^{-2}$~s$^{-1}$ or 1.92\% of the Ly$\alpha$ flux. 
Thus the Lyman continuum flux is 2.7 times brighter than the Lyman series 
lines (see Table~4).

We also plot in Figure~4 the Lyman continuum flux for the \citet{Fontenla2013}
models at the same wavelengths. These fluxes are also well fit by straight 
lines in these semilog plots. Note that the observed quiet Sun (SIRS)
Lyman continuum fluxes are similar to those of Model 13x1 and that the slopes 
of the model data steepen with decreasing solar activity. Because of the short 
wavelengths in the Lyman continuum compared to chromospheric temperatures,
the slopes of the continuum flux vs. wavelength are well fit by Wien's 
approximation to the Planck function, and the color temperatures obtained from 
the slopes are a good measure of the temperature where the optically thick 
continuum is formed.
We include in Table~4 the total Lyman continuum fluxes, ratios to the
Ly$\alpha$ flux, and the color temperatures. The Lyman continuum flux,
ratio to Ly$\alpha$, and color temperature of the SIRS data set are all
similar to Model 1301.

Figure~5 shows a comparison of the ratios of fluxes in the 70--80~nm,
80--91.2~nm, and 91.2--117~nm wavelength bands to the Ly$\alpha$ flux. 
The solid lines are least-squares fits to the EUV/Ly$\alpha$ 
flux ratios for the \citet{Fontenla2013} models.
The Lyman continuum is the largest contributor to the 80--91.2~nm wavelength 
band, but bright emission lines of O~III, O~IV, and N~IV dominate the 
70--80~nm passband. The SIRS and {\em SEE} solar data points for the 
91.2--117~nm passband are in excellent agreement with the model predictions.
For the 80--91~nm and 70--80~nm wavelength bands, the SIRS data agree with the 
models better than the {\em SEE} data.

\subsection{\it The 10--60~nm Portion of the EUV Spectrum}

The spectrum below 60~nm includes a number of features formed in the 
chromosphere, including the He~I continuum visible between 45 and 50.4~nm
and emission lines of He~I (58.4 and 53.7~nm) and He~II (30.4 and 25.6~nm).
He~II 30.4~nm is the brightest emission line in the 10--91.2~nm region.
This line is formed primarily by collisional excitation in the chromosphere,
but a portion of the emission is recombination following photoionization of 
He$^+$ by coronal radiation \citep{Avrett1976}.
There are also a number of transition region lines of O~II, O~III, O~IV, and
N~III located in this spectral region. Beginning with the  
Mg~X (61.0--62.5~nm) and Si~XII (49.9--52.1~nm) multiplets, 
coronal emission lines
increasingly dominate the spectrum at shorter wavelengths. Thus the 
10--60~nm portion of the EUV spectrum of the quiet Sun is a combination
of emission from the chromosphere, transition region, and corona.

Figure~6 compares the EUV/Ly$\alpha$ flux ratios with the 
\citet{Fontenla2013} models and the SIRS and {\em SEE} solar reference data. 
The agreement of the SIRS and quiet Sun {\em SEE} data with the models is very 
good, but the active Sun {\em SEE} data are low compared to the models for 
the 50--60~nm and 60--70~nm wavelength bands.
In the absence of any stellar data for comparison with the solar ratios 
or models, we suggest using least-squares fits to the models with the 
parameters listed in Table~5.

At wavelengths shorter than about 40~nm, modest interstellar absorption to the
nearest stars permits detection of EUV radiation, thereby providing empirical
tests of the accuracy with which the solar data and models can provide
estimates of the EUV flux ratios for different spectral type stars.
We compare in Figure~7 the solar 10--20~nm flux ratios to {\em EUVE} flux 
ratios (corrected for interstellar absorption) $R_{\rm ISM}$ versus
scaled Ly$\alpha$ fluxes for
15 stars with spectral types between F5~IV-V (Procyon) and M5.5~V
(Proxima Centauri). The least-squares fit to the data for the F5--K7 
stars is log[f(10--20~nm)/f(Ly$\alpha$)] = $-1.357 \pm 0.127$ + 
$0.344 \pm 0.094$ log[f(Ly$\alpha$)]. The mean deviation about this fit 
line is 20.5\% and the rms deviation is 29.0\%. There is
no apparent trend with spectral type. 
Excluding the EV~Lac and AU~Mic flare data (see later in this section), the 
remaining five M stars have nearly the same EUV flux ratios. Since the 
linear coefficient in the least-squares fit to these data is smaller than its 
uncertainty, we fit the data with a constant value,
log[f(10--20~nm)/f(Ly$\alpha$)] = --0.491.
The M star flux ratios have a small mean deviation of 12.6\% and a small 
rms deviation of 15.1\%.
Also plotted in Figure~7 are flux ratios
for the \citet{Fontenla2013} models. The flux ratios of $\alpha$~Cen A+B 
and Procyon are very
close to the quiet Sun models and data, and the flux ratios for the other
G- and K-type stars are also consistent with the solar model ratios. 
Our recommend fitting relations are summarized in Table~5.

Figure 8 is a similar comparison of solar and stellar data and solar model flux
ratios for the 20--30~nm wavelength region.
The least-squares fit to the data for the F5--K7 
stars is log[f(10--20~nm)/f(Ly$\alpha$)] = $-1.300 \pm 0.224$ + 
$0.309 \pm 0.164$ log[f(Ly$\alpha$)] with a mean deviation of 
47.6\% and an rms deviation of 56.6\%. Again excluding the EV~Lac and
AU~Mic flare data, the five M stars can be fit by
log[f(20--30~nm)/f(Ly$\alpha$)] = --0.548. The mean deviation about this 
fit is 24.3\% and the rms deviation is 26.9\%. 
The solar model ratios are consistent with the 
$\alpha$~Cen~A+B and the other F-, G-, and K-star data.

Finally, Figure 9 shows a similar plot for the 30--40~nm data. Since 
the interstellar absorption corrections exceed a factor of 3 for this 
bandpass when log [N(H~I)] $>18.2$, the flux ratios for AU~Mic and AD~Leo
depend critically on the uncertainties in the N(H~I) parameter. The models fit 
the solar data well, but have a slightly different slope than the 
F5~V--K7~V stars. Since the linear coefficient in the least-squares fit to
the F5--K7 stellar flux ratios is consistent with zero, we fit these data
with log[f(30--40~nm)/f(Ly$\alpha$)] = --0.882, with a mean deviation of 
37.1\%  and an rms deviation of 41.0\%. The M star flux ratios, except for the 
two flaring stars, can be fit with log[f(30--40~nm)/f(Ly$\alpha$)] = 
--0.602 with a mean deviation of 18.9\% and an rms deviation of 20.5\%.
Table~5 summarizes these fits and the dispersions of the stellar data 
points about these fits.

Since we do not have Ly$\alpha$ fluxes for EV~Lac and AU~Mic during their 
flares, we consider two different ways of representing their flare flux ratios.
In Figures 7--9, we plot the ratios of the observed EUV flare fluxes divided 
by the reconstructed quiesent Ly$\alpha$ fluxes vs. 
the scaled quiescent Ly$\alpha$ fluxes. This almost certainly 
overestimates the flux ratios and places the data points at unrealistically low
scaled Ly$\alpha$ flux levels. The dashed lines extending downwards
and to the right in the figures show the location of the ratios with increasing
Ly$\alpha$ flux. The correct ratios should lie along these dashed lines.
We estimate the most likely values of the Ly$\alpha$ flux during the flares
by noting the factors by which the EUV fluxes during the flare of AU Mic 
exceed the quiescent values and using the formulae in Table 5. The symbols at
the lower right end of the dashed lines in the figures indicate the most likely
values for the flare ratios and scaled Ly$\alpha$ fluxes for the two 
stars. These flare ratios are close to 
the mean nonflare ratios for M dwarf stars, indicating that the  
fits obtained using the quiescent M star data should be useful for estimating
EUV fluxes of M dwarfs over a wide range of activity, provided the 
Ly$\alpha$ flux is appropriate for the given level of activity. 

\subsection{\it Errors in the Flux Ratio Estimates}

There are three sources of error in our technique: errors in the EUV fluxes,
errors in the reconstructed Ly$\alpha$ fluxes, and errors associated with 
stellar variability since the EUV and Ly$\alpha$ fluxes were not obtained
at the same time. The mean and rms dispersions about the fit lines reflect all
three sources of error. The uncertainties in the Ly$\alpha$ reconstructions
are probably in the range 10--30\%, depending on the quality of the 
observations and the complexity of the interstellar medium velocity structure.
The dispersions for the F5--K5 stars about the fits lie in the range 20--37\%. 
This is consistent with errors in the EUV fluxes and errors associated with 
stellar variability each being in the range of 10--20\%. The dispersions 
for the M stars are smaller, 13--24\%. 
Although this was unexpected, it may result from the 
exclusion of obvious flaring events during the measurements of the 
Ly$\alpha$ and EUV fluxes. Also the M stars are located closer to the Sun
than the F5--K5 stars, in which case the velocity structure of the ISM 
should be simpler and the Lyman$\alpha$ reconstructions more reliable. 
Expansion of the M star data set would be helpful in understanding the relative
contributions of the three components to the dispersions.

\section{DISCUSSION}

\citet{Sanz-Forcada2011} (hereafter SF2011) 
developed a technique for predicting EUV fluxes
based on an emission measure analysis of observed stellar X-ray spectra. This
method predicts the emission line spectrum between 0.1 and 91.2~nm but may not
accurately include the Lyman continuum, important between 70 and 91.2~nm, 
or the He~I
and He~II continuua below 50.4~nm and 22.8~nm. In Table~6, we compare predicted
EUV flux ratios for the five stars in SF2011 Table~4 for which 
\citet{Linsky2013} list Ly$\alpha$ fluxes.

We checked {\em MAST} to find that only one star, $\epsilon$~Eri, 
listed in Table~4 of SF2011 was observed spectroscopically by {\em EUVE}. 
In the 10--20~nm band, the flux ratio predicted using our formula
(see Table~6) and SF2011 are both consistent with the {\em EUVE} spectrum of 
$\epsilon$~Eri corrected for interstellar absorption, $R_{\rm ISM}$. 
In the 20--30~nm band, our formula predicts a flux ratio 0.22 dex below the 
{\em EUVE} data and the SF2011 prediction is 0.09 dex larger than the 
{\em EUVE} data. In the 30--40~nm band, the SF2011 prediction is 0.07 dex 
larger than observed by {\em EUVE}, and our formulae predicts a flux ratio 
0.14 dex smaller. In the 70--91.2~nm band, where the Lyman 
continuum is an important contributor to the emission, the inclusion of the 
Lyman continuum flux would likely place the SF2011 ratio about 0.3 dex 
above our model prediction for $\epsilon$~Eri. 

The four other stars listed in Table~6 without {\em EUVE} spectroscopic data 
show no clear pattern
in the predicted flux ratios based on our formulae and SF2011. For HD~209458 
(G0~V), the upper limits predicted by SF2011 all lie below, and at some 
wavelengths far below, the flux ratios predicted by our formulae. This 
highlights
the problem of computing emission measure distributions based on very weak
or upper limits to the X-ray fluxes. For HD~189733 (K1~V), the flux ratios 
predicted by SF2011 are systematically high compared to our formulae,
and for GJ~436 (M3~V) they are 
systematically low compared to our formulae. For GJ~876 (M5.0~V) the SF2011
flux ratios are very low compared to our formulae in the 10--20 and 20--30~nm 
bands, but comparable in the longer wavelength bands.

Finally, we compare in Table~6 the 91.2--117~nm flux ratios predicted by our 
formulae with the fluxes obtained using the emission measure 
analysis technique listed in the X-exoplanets website
\footnote{http://sdc.cab.inta-csic.es/xexoplanets/jsp/exoplanetsform.jsp} 
described in SF2011 divided by the reconstructed Ly$\alpha$ fluxes 
\citep{Linsky2013}. The flux ratios computed from the predicted fluxes in
the X-exoplanets website are 0.6--2.5~dex below those predicted by our 
formulae.
The excellent $\epsilon$~Eri {\em FUSE} data provide a clear test of the two
prediction methods. The f(91.2--117~nm)/f(Ly$\alpha$) ratio obtained from
the {\em FUSE} emission line fluxes \citep{Redfield2002} 
and estimated Lyman series flux is 0.2 dex below that 
predicted by our method but 0.9 dex above that obtained from the data 
in the X-exoplanets website. The missing flux in the X-exoplanet website 
predictions likely indicates the inadequate treatment of emission lines formed 
at temperatures below $2\times 10^5$~K.

\section{CONCLUSIONS}

EUV fluxes from host stars control the photochemistry,
heating, and mass loss from the outer atmospheres of exoplanets, especially 
for those exoplanets located close to their host stars. 
The objective of this study is to develop a useful technique for predicting
the EUV fluxes of F5--M5 dwarf stars, since there are only a few measurements 
of stellar EUV fluxes and interstellar absorption prevents measurements  
between 40 and 91.2~nm for all nearby dwarfs stars except for the Sun. 
Our technique employs ratios of EUV fluxes in wavelength bands to the 
Ly$\alpha$ flux, because models of the 
solar chromosphere, transition region, and corona show that the EUV flux 
scales with the Ly$\alpha$ flux. Moreover, models of solar regions 
with different amounts of magnetic heating show temperature structures
with similar shapes but displaced deeper into the atmosphere (and thus 
higher densities) with increasing magnetic heating. These empirical and 
theoretical arguments gives us confidence that the
ratios of the EUV flux in various wavelength bands to the Ly$\alpha$ flux 
should vary smoothly with stellar activity at least for stars that do not 
differ too greatly from the Sun in spectral type. Models of stellar 
chromospheres and transition regions comparable in detail with the solar 
models of \citet{Fontenla2013} are needed
to confirm the range of stars for which our technique is useful. Until such
models are available, observations of the few stars in the 91.2--117.0~nm
range by the {\em FUSE} satellite and in the 10--40~nm range by the {\em EUVE} 
satellite show that our ratio technique is useful for F5--M5 dwarfs.

Table 5 summarizes our recommended formulae for predicting the
log [f($\Delta\lambda$)/f(Ly$\alpha$)]
ratios in nine wavelength bands. 
For the 10--20~nm, 20--30~nm, and 30--40~nm wavelength bands, we recommend
using the formulae based on the stellar fluxes observed by {\em EUVE}.
Our formulae predict flux ratios similar to the observations and to
the emission measure analysis predictions of SF2011, as indicated by comparison
with the {\em EUVE} data for $\epsilon$~Eri.
For the 40--91.2~nm wavelength range, where there are no reliable stellar 
observations to compare with the solar fluxes or models, 
we suggest using the formulae based on the \citet{Fontenla2013} solar models 
to predict flux ratios for F7--K7 dwarf stars. 
For M stars, we suggest adding 0.2 dex to the solar 
ratios, the mean displacement of the M stars from the warmer stars in the 
20--30 and 30--40~nm bands. In the 91.2--117~nm band, the agreement between
the {\em FUSE} data and our model predictions shown in 
Figure~3 suggests that the ratios for M stars may be the same as for the 
warmer stars. Fits to the flux ratios based on the {\em FUSE} data and our 
formulae are in good agreement. On the other hand, the predictions of the 
emission measure analysis models in the X-Exoplanets website are far below
our models and the {\em FUSE} observations of $\epsilon$~Eri.
Figures 7--9 and Table~5 show that for the F5--K5 stars the mean deviations
from the fit lines lie in the range 20--48\%,
and for the nonflaring M stars the mean deviations lie in the range 12--24\%.
Thus our formulae should be useful in predicting the EUV flux ratios for 
F5--M5 dwarf stars. We note that the flux ratios based on the 
\citet{Fontenla2013} solar models closely match the solar data, as expected, 
but they also come remarkably close to matching the stellar flux ratios.

\acknowledgements

This work is supported by NASA through grants NNX08AC146, NAS5-98043, 
and HST-GO-11687.01-A to the University of Colorado at Boulder. 
KF acknowledges support through the NASA Nancy Grace Roman Fellowship
during a portion of this work. We thank Jorge Sanz-Forcada for calling 
attention to an error in our reduction of the {\em EUVE} data,
Martin Snow and Tom Woods for providing the solar irradiance data, 
and Jurgen Schmitt for
providing X-ray luminosities for M dwarf stars. We appreciate the availability
of {\em HST} data through the MAST website hosted by the Space Telescope 
Science Institute, stellar data though the SIMBAD database operated at
CDS, Strasbourg, France,
and the X-exoplanets Archive at the CAB \citep{Sanz-Forcada2011}.
Finally, we thank the referee for his insightful and very useful comments.
{\it Facilities:} \facility{FUSE}, \facility{EUVE}.

\begin{figure}
\includegraphics{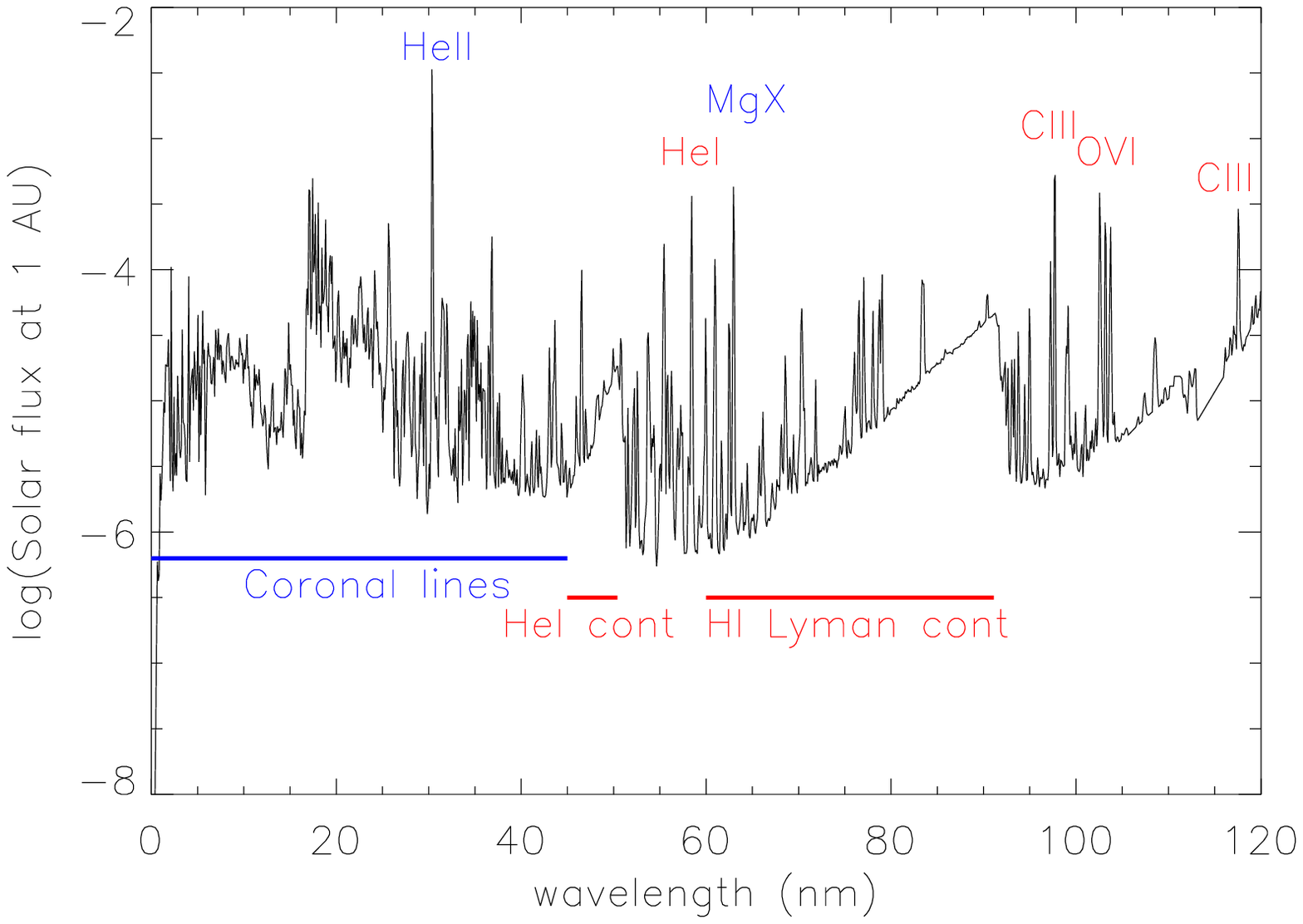}
\caption{The Solar Irradiance Reference Spectrum (SIRS) obtained at 
solar minimum (March--April 2008). Flux units are Wm$^{-2}$nm$^{-1}$ at 1 AU.
Important emission lines and continua are identified.}
\end{figure}
 
\begin{figure}
\includegraphics{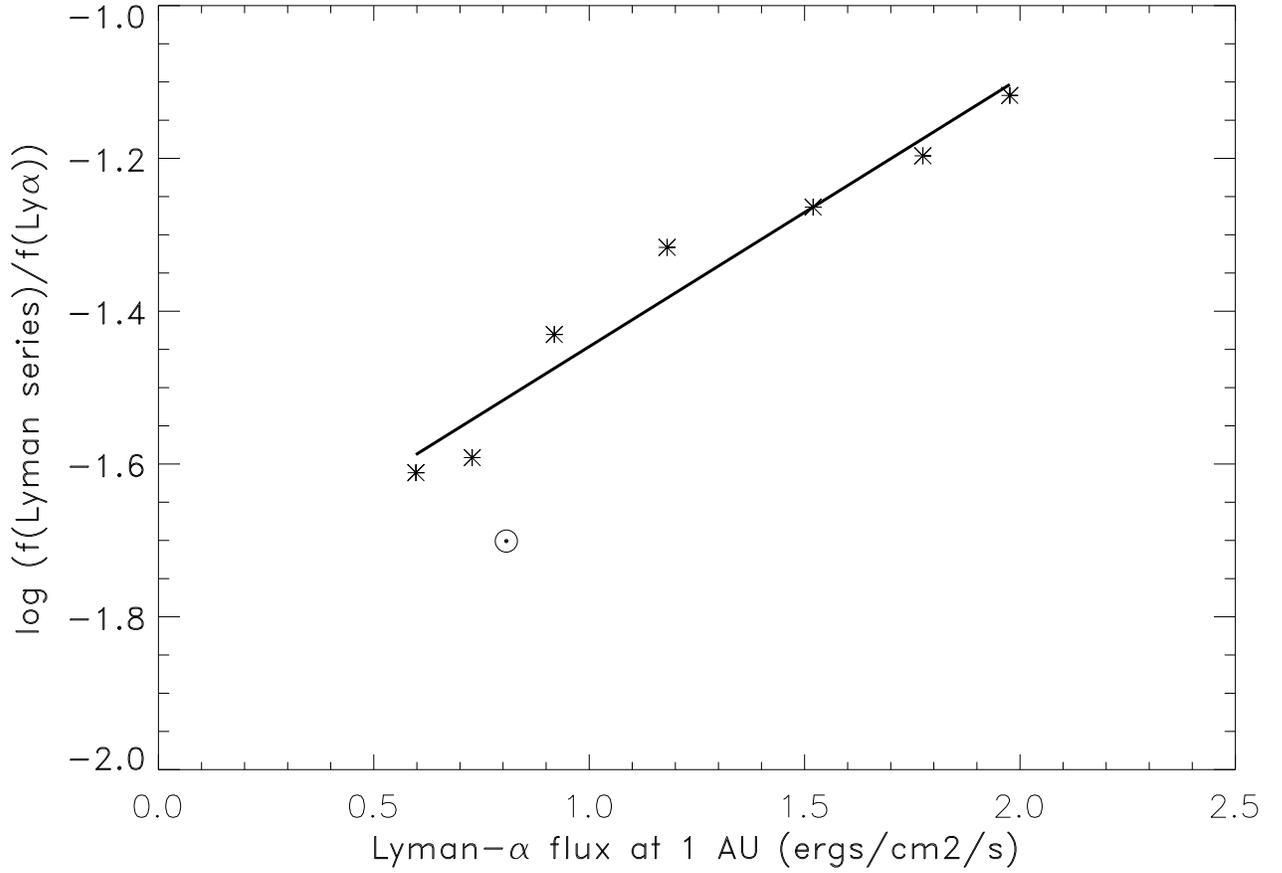}
\caption{Ratio of the Lyman series flux from Ly$\beta$ to 91.2~nm divided 
by the Ly$\alpha$ flux for the \citet{Fontenla2013} semiempirical models
13x0 to 13x8 (asterix symbols from left to right). The solid line is a 
least-squares fit to the data points. The SIRS
flux ratio is represented by the solar symbol.}
\end{figure}

\begin{figure}
\includegraphics{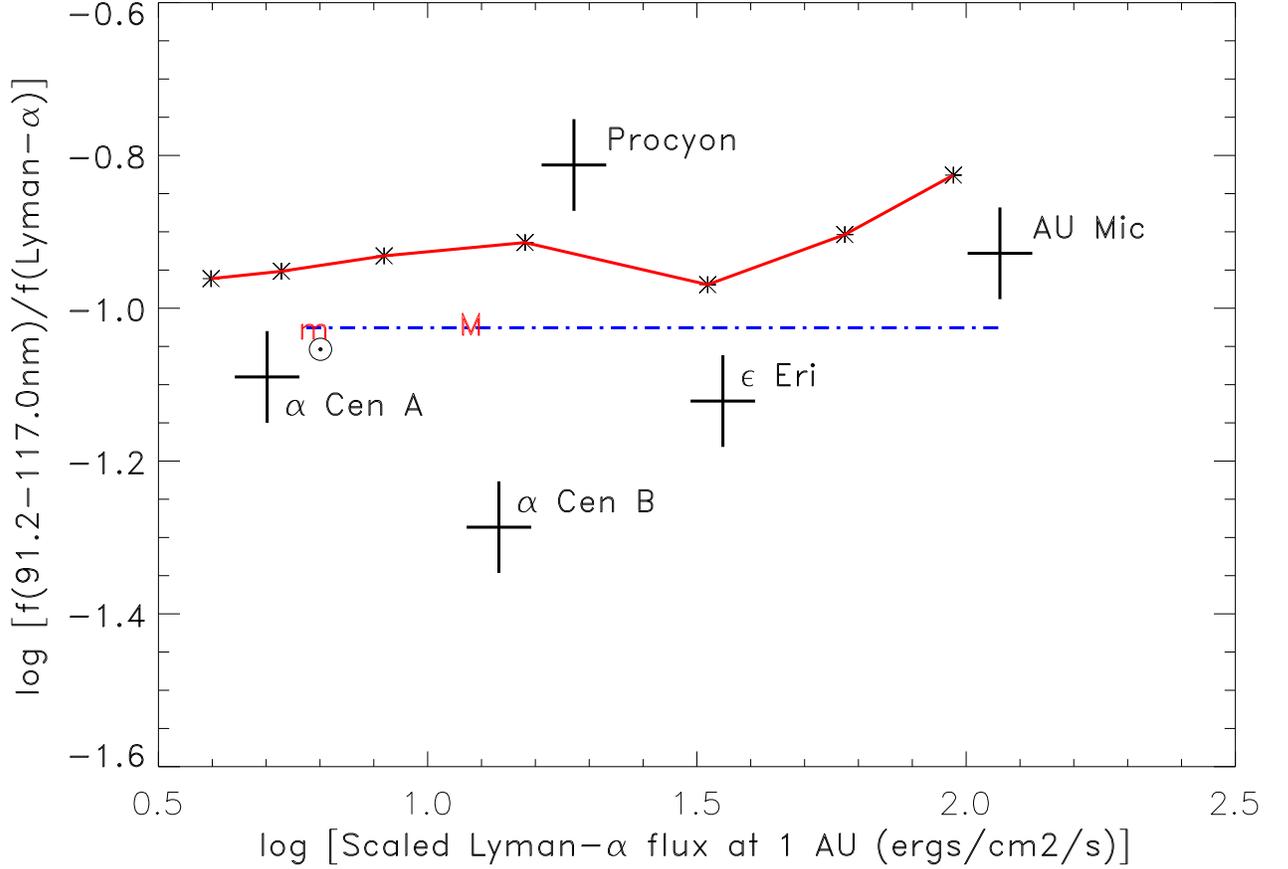}
\caption{Ratios of the total flux between 91.2 and 117.0 nm divided by the
Ly$\alpha$ flux at 1~AU scaled by the ratio of stellar radii, 
$(R_{Sun}/R_{star})^2$. The solid-line-connected asteriskes (red line) 
are the total flux in 
this passband for the \citet{Fontenla2013} semi-empirical models 13x0 to 13x8
(from left to right).
Flux ratios for five stars based on {\em FUSE} spectra and 
reconstructed Lyman series fluxes are shown as $\pm 15$\% error bar symbols. 
The Sun symbol is the ratio for the SIRS quiet Sun data set. The dash-dot 
(blue) line is the least-squares fit to the stellar and SIRS data. 
The ``m'' and ``M''
symbols are the solar minimum and maximum data obtained with the {\em SEE} 
instrument on the {\em TIMED} spacecraft \citep{Woods2005}.}
\end{figure}

\begin{figure}
\includegraphics{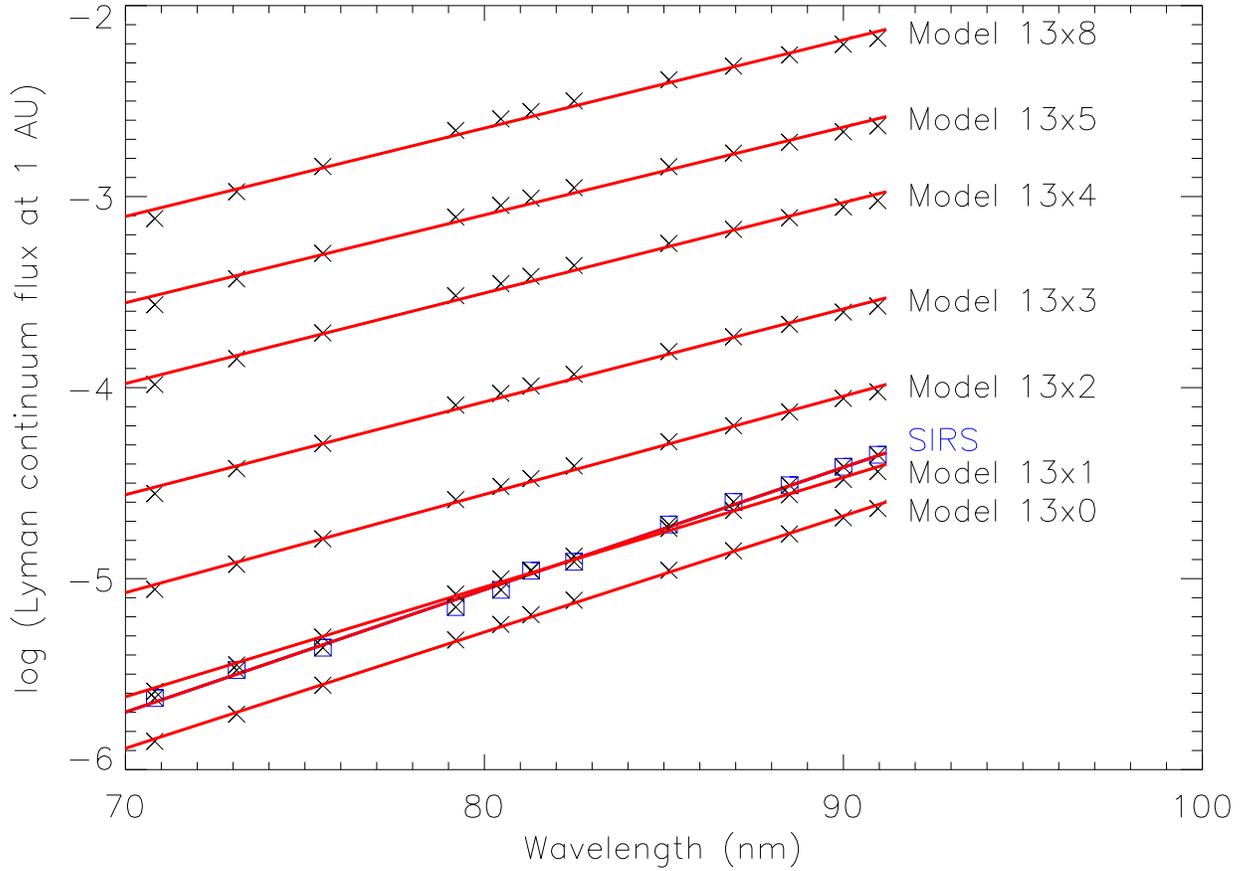}
\caption{Plots of the Lyman continuum flux (Wm$^{-2}$nm$^{-1}$)
at a distance of 1~AU for the 
quiet Sun SIRS data set and for the semiempirical solar irradiance 
models of \citet{Fontenla2013}. X-symbols are fluxes 
for spectral regions with no apparent emission lines. Box symbols are for the
SIRS data. Solid red lines are least-squares fits to the data.}
\end{figure}

\begin{figure}
\includegraphics{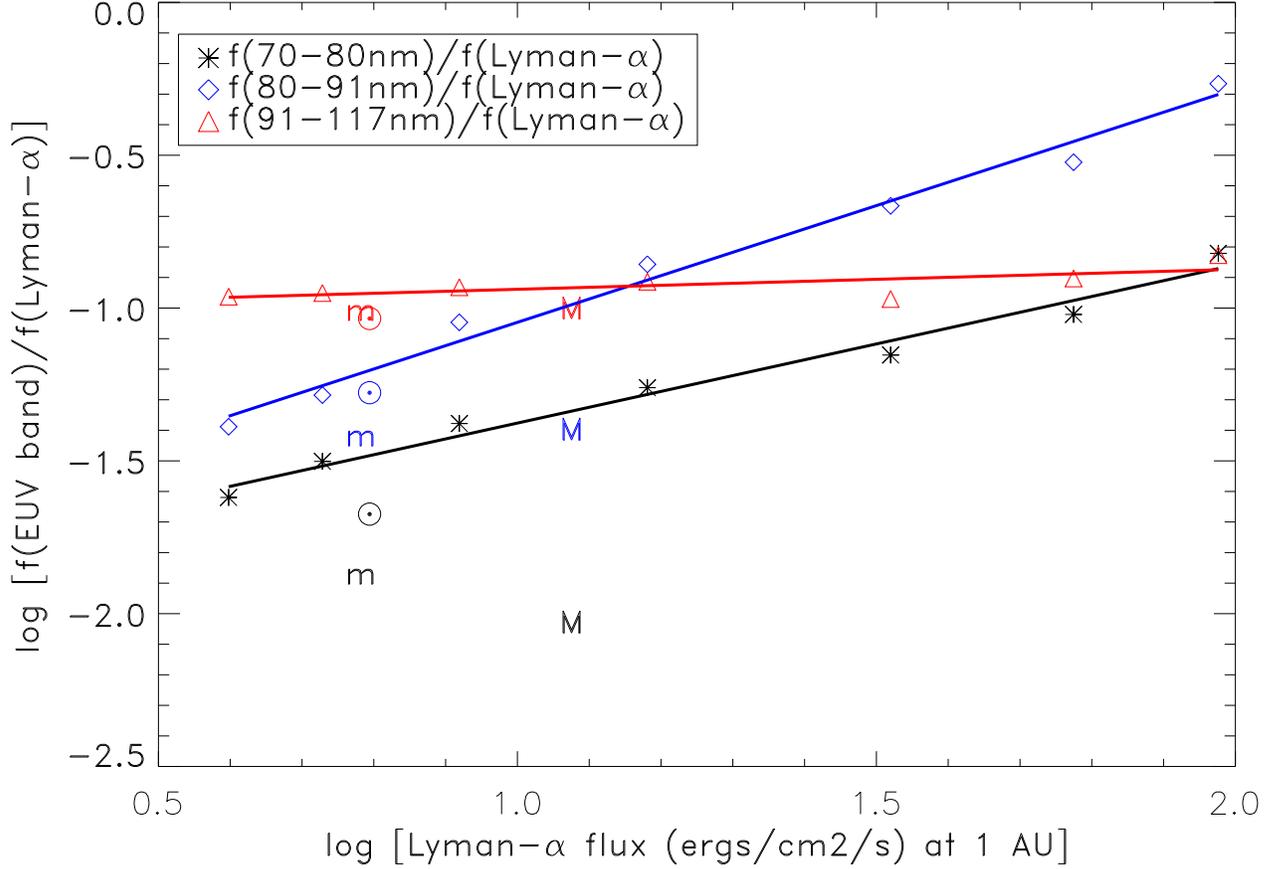}
\caption{Plots of the ratios of the fluxes in the 70--80~nm, 80--91.2~nm,
and 91.2--117~nm wavelength bands divided by the Ly$\alpha$ flux. 
The symbols are total fluxes in each wavelength band for 
\citet{Fontenla2013} models 13x0 to 13x8 (left to right). The solid lines 
are least-squares fits to each data set. 
The Sun symbols are for the SIRS quiet Sun fluxes in these wavelength bands,
and the ``m'' and ``M'' symbols are the solar minimum and maximum data
obtained with the {\em SEE} experiment on the {\em TIMED} spacecraft for the 
wavelength bands.
The 80--91.2~nm passband is mainly Lyman continuum flux, whereas the 70--80~nm
passband contains strong transition-region lines of O~III, O~IV, and N~IV
that are likely overestimated in the models.}
\end{figure}

\begin{figure}
\includegraphics{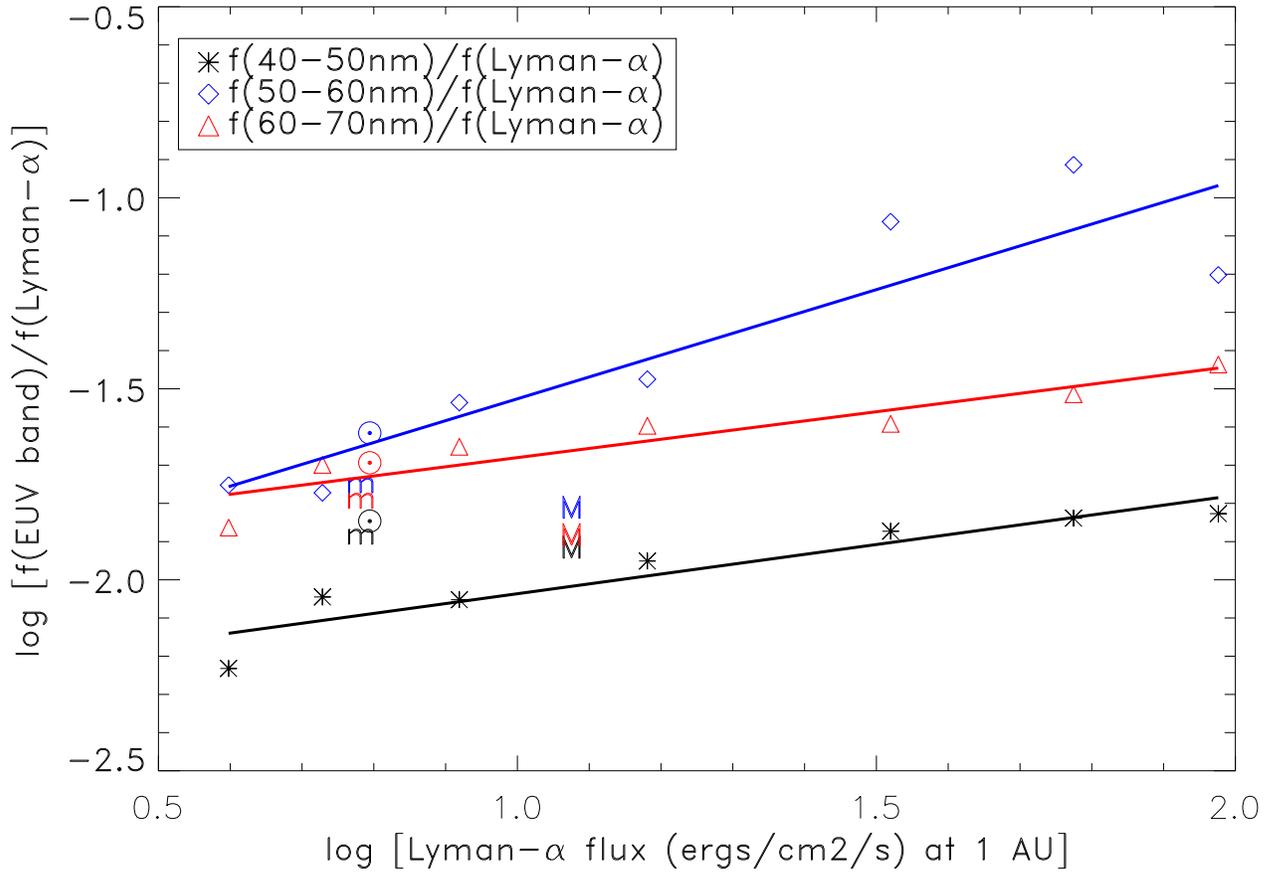}
\caption{Same as Figure 5 except for the 40--50~nm, 50--60~nm, and 60--70~nm
wavelength bands.}
\end{figure} 

\begin{figure}
\includegraphics{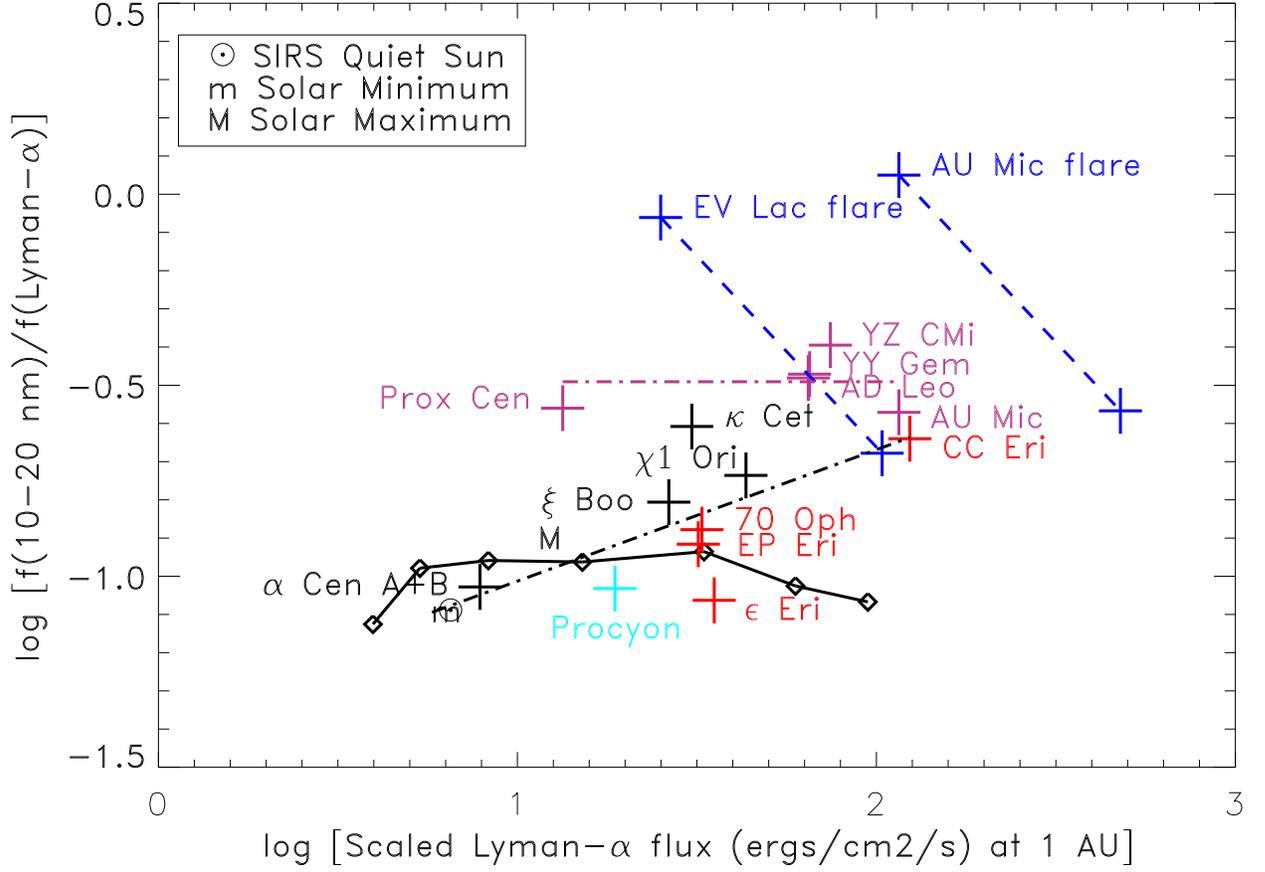}
\caption{Ratios of the intrinsic flux between 10 and 20~nm (corrected for 
interstellar absorption) divided by the reconstructed 
Ly$\alpha$  flux vs. the reconstructed Ly$\alpha$ flux at 1 AU
scaled by the ratio of stellar radii, $(R_{Sun}/R_{star})^2$.
The solid line-connected diamonds are the total flux ratios in this 
passband for the \citet{Fontenla2013} semiempirical models 13x0 to 13x8
(from left to right). Flux ratios for one F star (cyan), four G stars 
(black), four K stars (red), and five M stars (plum) based on {\em EUVE} 
spectra are shown as $\pm 15$ error bar symbols. The dash-dot 
(black) line is the least-squares fit to the solar and F, G, and K star ratios.
The plum dash-dot line is the mean of the M star ratios excluding
the EV Lac flare and AU Mic flare data. Flux ratios for EV Lac 
and AU Mic during flares (blue) are plotted two ways. The upper left symbols 
are ratios of EUV flare fluxes to quiescent Ly$\alpha$ fluxes. Dashed lines
extending to the lower right indicate the ratios for increasing Ly$\alpha$
flux. The symbols at the lower end of the dashed lines are ratios obtained 
using the most likely values of the Ly$\alpha$ fluxes during flares
(see text).
The ``m'' and ``M'' symbols are the solar minimum and maximum data obtained 
with the SEE instrument on the {\em TIMED} spacecraft \citep{Woods2005}.  
The Sun symbol is the ratio for the SIRS quiet Sun data set.}
\end{figure} 

\begin{figure}
\includegraphics{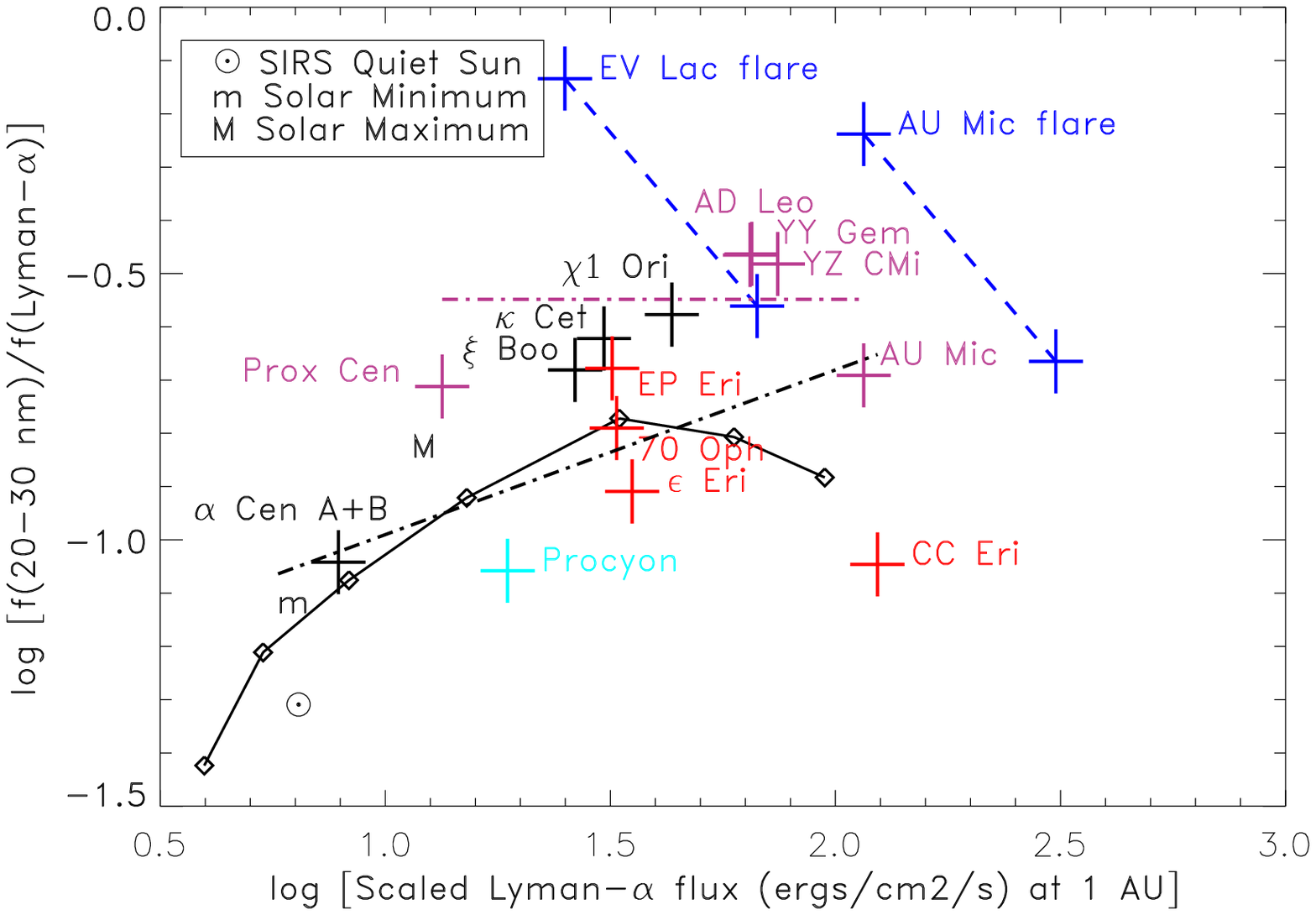}
\caption{Same as Figure 7 except for the 20--30~nm wavelength interval.}
\end{figure} 

\begin{figure}
\includegraphics{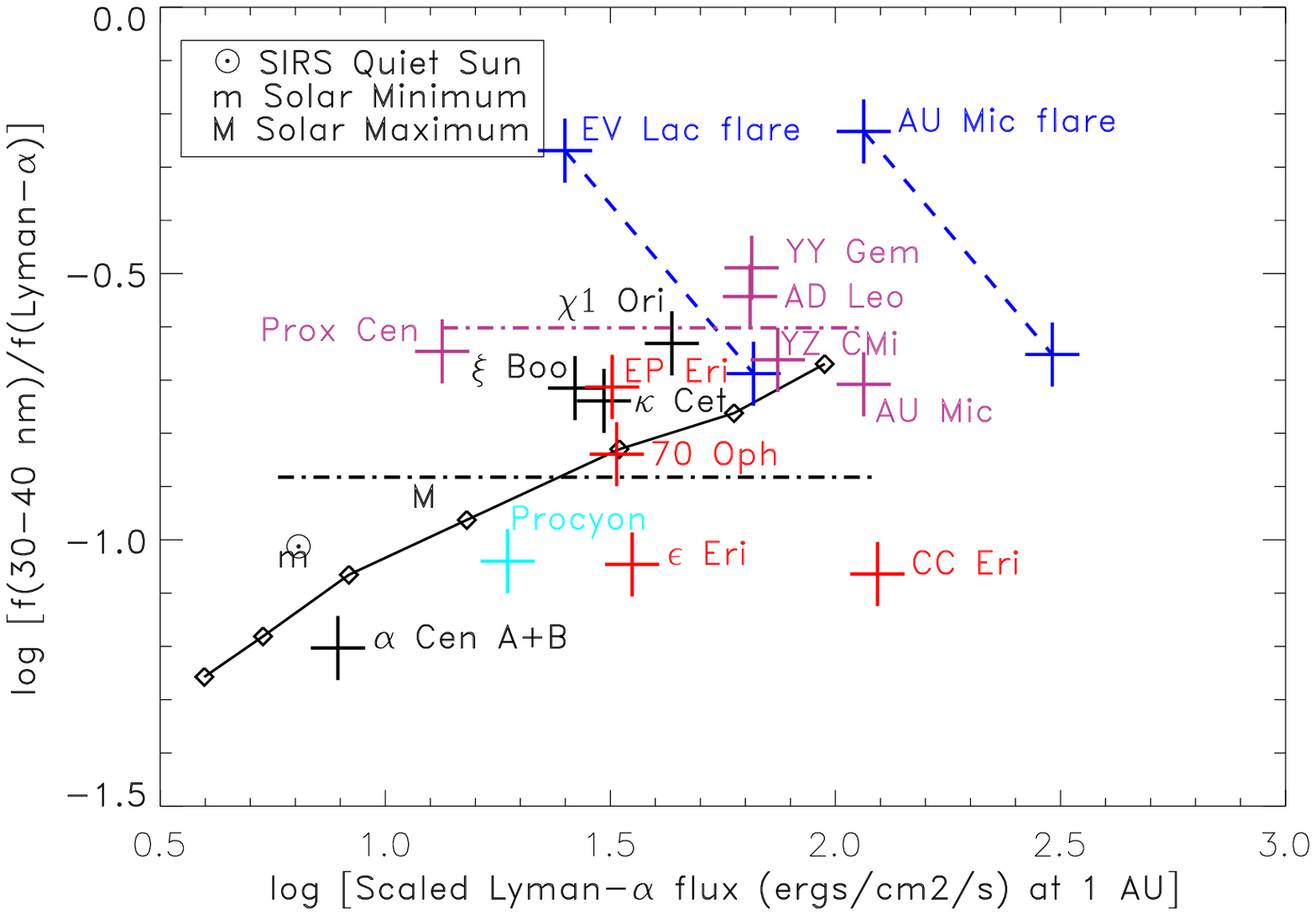}
\caption{Same as Figure 7 except for the 30--40~nm wavelength interval.}
\end{figure} 

\begin{deluxetable}{lcccccc}
\tablewidth{0pt}
\tablenum{1}
\tablecaption{\bf Solar EUV Fluxes (erg cm$^{-2}$ s${-1}$) in Wavelength Bands}
\tablehead{\colhead{Wavelength} &  \multicolumn{2}{c}{SIRS (Solar Minimum)} &
\multicolumn{2}{c}{SEE (Solar Minimum)} & \multicolumn{2}{c}{Solar Maximum}\\
\colhead{Band (nm)} & \colhead{f(EUV)} & \colhead{f(EUV)/f(Ly$\alpha$)} &
\colhead{f(EUV)} & \colhead{f(EUV)/f(Ly$\alpha$)} &
\colhead{f(EUV)} & \colhead{f(EUV)/f(Ly$\alpha$)}}
\startdata

Ly$\alpha$    &  5.95 &        & 5.78  &        & 11.5  &      \\
10--20nm      & 0.451 & 0.0758 & 0.440 & 0.0761 & 1.35  & 0.118\\
20--30nm      & 0.276 & 0.0465 & 0.422 & 0.0730 & 1.64  & 0.143\\
30--40nm      & 0.548 & 0.0921 & 0.514 & 0.0889 & 1.33  & 0.115\\
40--50nm      & 0.0788& 0.0132 & 0.0718& 0.0124 & 0.316 & 0.0144\\
50--60nm      & 0.134 & 0.0225 & 0.0977& 0.0169 & 0.166 & 0.0145\\
60--70nm      & 0.112 & 0.0188 & 0.0890& 0.0154 & 0.141 & 0.0123\\
70--80nm      & 0.115 & 0.0193 & 0.0721& 0.0125 & 0.0995& 0.00865\\
80--91.2nm    & 0.287 & 0.0483 & 0.204 & 0.0354 & 0.426 & 0.0370\\
91.2--117nm   & 0.502 & 0.0844 & 0.527 & 0.0911 & 1.060 & 0.0922\\
117--130nm-Ly$\alpha$ & &      & 0.538 & 0.0930 & 0.779 & 0.0677\\ 
130--140nm    &       &        & 0.543 & 0.0939 & 0.811 & 0.0705\\
140--150nm    &       &        & 0.558 & 0.0965 & 0.689 & 0.0599\\
150--160nm    &       &        & 1.367 & 0.237  & 1.634 & 0.142\\
160--170nm    &       &        & 3.174 & 0.549  & 3.631 & 0.316\\
170--180nm    &       &        & 9.831 & 1.701  & 11.23 & 0.977\\
\enddata
\end{deluxetable}

\begin{deluxetable}{lcccccc}
\tablewidth{0pt}
\tablenum{2}
\tablecaption{\bf Stellar Fluxes (ergs cm$^{-2}$ s$^{-1}$) at 1 AU in 
Different Wavelength Bands}
\tablehead{\colhead{Parameter} &  \colhead{Procyon} & \colhead{SIRS} &
\colhead{$\alpha$~Cen~A} & \colhead{$\alpha$~Cen~B} & 
\colhead{$\epsilon$~Eri} & \colhead{AU Mic}}
\startdata
Spectral Type     & F5 IV-V& G2 V & G2 V  & K0 V  & K1 V  & M0 V\\
d(pc)             & 3.50   &      & 1.325 & 1.255 & 3.216 & 9.91\\
Age(Gyr)\tablenotemark{a} & 1.85 & 4.566 & $4.4\pm0.5$ & $4.4\pm0.5$ & 0.43 &
$0.020\pm0.010$\\
f(Ly$\alpha$)     & 77.1   & 5.95 & 7.54  & 10.1  & 21.5  & 43.0\\
f({\em FUSE} data without Lyman lines) & 6.46   &      & 0.374 & 0.168 & 
0.650 & 2.61\\
f(Lyman series)   & 5.41   & 0.242& 0.239 & 0.354 & 0.976 & 2.47\\
f(91.2--117.0 nm) & 11.87  & 0.507& 0.613 & 0.522 & 1.626 & 5.08\\
f(91.2--117.0 nm)/f(Ly$\alpha$) & 0.154 & 0.0852 & 0.0813 & 0.0517 & 0.0756 & 
0.118\\
\enddata
\tablenotetext{a}{Stellar age references: Procyon \citep{Takeda2007},
Sun \citep{Allegre1995}, $\alpha$~Cen~A, $\alpha$~Cen~B, and $\epsilon$~Eri
\citep{Barnes2007}, and AU Mic \citep{Zuckerman2001}.}
\end{deluxetable}

\begin{deluxetable}{lcccccccccc}
\setlength{\tabcolsep}{1mm}
\tabletypesize{\footnotesize}
\rotate
\tablewidth{0pt}
\tablenum{3}
\tablecaption{\bf Ratios of {\it EUVE} Fluxes in Wavelength Bands Divided 
by the Intrinsic Ly$\alpha$ Fluxes}
\tablehead{\colhead{Star} & \colhead{$R_{\star}/R_{\odot}$} &
\colhead{EUVE ID} & 
\colhead{f(Ly$\alpha$)\tablenotemark{a}} & \colhead{log[N(HI)]} &
\multicolumn{2}{c}{10--20 nm} & \multicolumn{2}{c}{20--30 nm} & 
\multicolumn{2}{c}{30--40 nm}\\
 & & \colhead{(ks)} & & & 
\colhead{log$R$} & \colhead{log$R_{\rm ISM}$} &
\colhead{log$R$} & \colhead{log$R_{\rm ISM}$} &
\colhead{log$R$} & \colhead{log$R_{\rm ISM}$}}
\startdata

Procyon (F5 IV-V) & 2.03 & procyon\_\_9403122334N & 77.1 & 18.06 & 
--1.100 & --1.032 & --1.223 & --1.058 & --1.376 & --1.040\\
$\chi^1$ Ori (G0 V) & 0.98 & chi1\_ori\_\_9301261159N & 41.6 & 17.93 &
--0.780 & --0.736 & --0.713 & --0.577 & --0.890 & --0.631\\
 
$\alpha$ Cen (G2 V+K0 V) & 1.50 & alpha\_cen\_9703100800N & 17.64 & 17.61 &
--1.052 & --1.028 & --1.100 & --1.042 & --1.316 & --1.203\\
$\kappa$ Cet (G5 V) & 0.99 & kappa\_cet\_\_9510061036N & 30.0 & 17.89 &
--0.645 & --0.608 & --0.738 & --0.622 & --0.985 & --0.739\\
$\xi$ Boo (G8 V+K4 V) & 1.16 & xi\_boo\_\_9704200202N & 35.3 & 17.92 &
--0.849 & --0.806 & --0.807 & --0.681 & --0.958 & --0.715\\

70 Oph (K0 V+K4 V) & 1.13 & gj\_702\_\_9307021144N & 23.6 & 18.06 &
--0.942 & --0.878 & --0.960 & --0.790 & --1.190 & --0.839\\

$\epsilon$ Eri (K1 V) & 0.78 & eps\_eri\_\_9509051851N & 21.5 & 17.88 &
--1.104 & --1.063 & --1.024 & --0.909 & --1.270 & --1.046\\ 

EP Eri (K2 V) & 0.93 & gj\_117\_\_9412020500N & 27.6 & 18.05 &
--0.976 & --0.916 & --0.853 & --0.678 & --1.051 & --0.713\\

CC Eri (K7 V) & 0.66 & cc\_eri\_\_9509130049N & (54) & (18.1) &
--0.690 & --0.640 & --1.211 & --1.046 & --1.433 & --1.064\\

AU Mic flare (M0 V) & 0.61 & au\_mic\_\_9207141227N & 43.0 & 18.36 &
--0.034 & +0.050  & --0.564 & --0.238 & --0.894 & --0.233\\

AU Mic (M0 V) & 0.61 & Mean\tablenotemark{b} & 43.0 & 18.36 &
--0.663 & --0.571 & --1.068 & --0.691 & --1.362 & --0.708\\

YY Gem (dM1e+dM1e) & 0.88 & yy\_gem\_\_9502201531N\_1 & (50.0) & (18.0) &
--0.511 & --0.471 & --0.614 & --0.463 & --0.784 & --0.489\\

EV Lac flare (M3.5 V) & 0.35 & ev\_lac\_\_9309091718N & 3.07 & 17.97 &
--0.103 & --0.061 & --0.272 & --0.134 & --0.556 & --0.269\\

AD Leo (M3.5 V) & 0.38 & Mean\tablenotemark{c} & 9.33 & 18.47 &
--0.602 & --0.481 & --0.949 & --0.465 & --1.408 & --0.543\\

YZ CMi (M4.5 V) & 0.30 & Mean\tablenotemark{d} & 6.7 & (17.8) &
--0.421 & --0.395 & --0.584 & --0.482 & --0.848 & --0.662\\

Prox Cen (M5.5 V) & 0.15 & proxima\_cen\_\_9305211911N & 0.301 & 17.61 &
--0.580 & --0.560 & --0.775 & --0.712 & --0.771 & --0.646\\

\enddata
\tablenotetext{a}{Intrinsic Ly$\alpha$ flux (ergs cm$^{-2}$ s$^{-1}$) 
at a distance of 1 AU.}

\tablenotetext{b}{average of data sets au\_mic\_\_9307220306N and 
au\_mic\_\_9606121801N.}

\tablenotetext{c}{average of data sets ad\_leo\_\_9904092045,
ad\_leo\_\_9904251629N, ad\_leo\_\_9904050046N, ad\_leo\_\_9905061641N,
ad\_leo\_\_9303010544N, ad\_leo\_\_0003091327N, ad\_leo\_\_9605030109N,
and ad\_leo\_\_9904170332N.}

\tablenotetext{d}{average of data sets yz\_cmi\_\_9412210116N and
yz\_cmi\_\_9302250656N.}

\end{deluxetable}

\begin{deluxetable}{lccccccccc}
\rotate
\tablewidth{0pt}
\tablenum{4}
\tablecaption{\bf Lyman line and Lyman continuum fluxes 
(erg cm$^{-2}$ s$^{-1}$) at 1 AU\tablenotemark{a}}
\tablehead{\colhead{Line} &  \colhead{$\lambda$(nm)} & \colhead{SIRS} &
\colhead{f(1300)} & \colhead{f(1301)} & \colhead{f(1302)} & 
\colhead{f(1303)} & \colhead{f(1304)} & \colhead{f(1305)}
& \colhead{f(1308)}}
\startdata 
Ly$\beta$           &102.57& 0.0655 & 0.0422 & 0.0599 & 0.128  & 0.296 & 
0.670 & 1.302 & 2.094\\
Ly$\gamma$          &97.25 & 0.0155 & 0.0191 & 0.0267 & 0.0584 & 0.135 &
0.301 & 0.610 & 1.20\\
Ly$\delta$          &94.97 & 0.0081 & 0.0119 & 0.0165 & 0.0369 & 0.0863& 
0.206 & 0.420 & 0.844\\
Ly$\epsilon$        &93.78 & 0.00487& 0.00765& 0.0109 & 0.0244 & 0.0563 & 
0.137 & 0.279 & 0.546\\
Ly7                 &93.08 & 0.00323& 0.00496& 0.00692& 0.0160 & 0.0376 & 
0.0988 & 0.208 & 0.416\\
Ly8                 &92.62 & 0.00186& 0.00302& 0.00426& 0.0102 & 0.0251 & 
0.0709 & 0.157 & 0.327\\
Ly9\tablenotemark{b}&92.31 & 0.0010 & 0.0019 & 0.0025 & 0.0067 & 0.0175 & 
0.0556 & 0.128 & 0.279\\
Ly10                &92.10 &        &0.000738& 0.000781&0.00320& 0.0101 & 
0.0404 & 0.0978 & 0.232\\
Ly11+rest           &91.94 &        &0.00555& 0.00849 & 0.0238 & 0.0686 & 
0.226 & 0.583 & 1.288\\
Sum                 &      & 0.114  &0.0969 & 0.137  & 0.308 & 0.732 & 
1.80 & 3.78 & 7.22\\
Ly$\alpha$          &      & 5.95   & 3.96   & 5.35   & 8.30  & 15.17 & 
33.11  & 59.52 & 94.68\\
Sum/Ly$\alpha$      &      & 0.0192 & 0.0245 & 0.0256 & 0.0371 & 0.0483 &    
0.0545 & 0.0636 & 0.0763\\
Ly$\beta$/Ly$\alpha$&      & 0.0110 & 0.0107 & 0.0112 & 0.0155 & 0.0195 & 
0.0202 & 0.0219 & 0.0221\\
Ly$\alpha$/Ly$\beta$&      & 90.84  & 93.86  & 89.33  & 64.64  & 51.22 & 
49.45 & 45.70 & 45.22\\
f(91.2--117.0nm)    &      & 0.507  & 0.433  & 0.598 & 0.973 & 1.85 &
3.55 & 7.43 & 14.14\\
f(91.2--117.0)/f(Ly$\alpha$) & & 0.0852 & 0.109 & 0.112 & 0.117 & 0.122 &
0.107 & 0.125 & 0.149\\
Lycont              &      & 0.307 & 0.178 & 0.296 & 0.858 & 2.56 & 9.42 & 
23.9 & 68.2\\
Lycont/Ly$\alpha$   &      & 0.0516 & 0.0449 & 0.553 & 0.103 & 0.169 & 0.285 &
0.402 & 0.720\\
Ly$\alpha$/Lycont    &      & 19.4 & 22.2 & 18.1 & 9.67 & 5.93 & 3.51 & 2.49 &
1.39\\
T(color) (K)        &      & 12,210 & 12,640 & 13,230 & 14,360 & 14,930 &
15,160 & 15,480 & 15,390\\
\enddata
\tablenotetext{a}{Observed flux from the quiet Sun (SIRS) and 
Fontenla et al. (2013) semiempirical models 1300 to 1308.}
\tablenotetext{b}{Estimated Ly9 flux from the blended feature.}
\end{deluxetable}

\begin{deluxetable}{lccc}
\tablewidth{0pt}
\tablenum{5}
\setlength{\tabcolsep}{1mm}
\tablecaption{\bf Formulae for Estimating EUV Fluxes (erg cm$^{-2}$ s$^{-1}$) 
in Wavelength Bands}
\tablehead{\colhead{Wavelength} &  
\multicolumn{3}{c}{log[f($\Delta\lambda$/f(Ly$\alpha$)]}\\ 
\colhead{Band (nm)} & \colhead{F5--K7 V stars} & \colhead{M V stars} &
\colhead{F5--M5 V stars}}

\startdata

10--20nm (stars) & --1.357+0.344 log[f(Ly$\alpha$)] & --0.491 &\\
20--30nm (stars) & --1.300+0.309 log[f(Ly$\alpha$)] & --0.548 &\\
30--40nm (stars) & --0.882 & --0.602 &\\
40--50nm (models) & & & -2.294+0.258 log[f(Ly$\alpha$)]\\

50--60nm (models) & & & -2.098+0.572 log[f(Ly$\alpha$)]\\

60--70nm (models) & & & -1.920+0.240 log[f(Ly$\alpha$)]\\

70--80nm (models) & & & -1.894+0.518 log[f(Ly$\alpha$)]\\

80--91.2nm (models) & & & -1.811+0.764 log[f(Ly$\alpha$)]\\

91.2--117nm (models) & & & -1.004+0.065 log[f(Ly$\alpha$)]\\

Lyman series (models) & & & -1.798+0.351 log[f(Ly$\alpha$)]\\

91.2--117nm (stars) & & & --1.025\\
10--20nm mean deviation & 20.5\%  & 12.6\% &\\
10--20nm rms deviation  & 29.0\%  & 15.1\% &\\
20--30nm mean deviation & 47.6\%  & 24.3\% &\\
20--30nm rms deviation  & 56.6\%  & 26.9\% &\\
30--40nm mean deviation & 37.1\%  & 18.9\% &\\
30--40nm rms deviation  & 41.0\%  & 20.5\% &\\
91.2--117nm mean deviation & & & 29.5\%\\
91.2--117nm rms deviation & & & 35.0\%\\
\enddata
\end{deluxetable}

\begin{deluxetable}{lccccc}
\tablewidth{0pt}
\tablenum{6}
\tablecaption{\bf Comparison of EUV Flux Ratios 
log [f($\Delta\lambda$)/f(Ly-$\alpha$)]}
\tablehead{\colhead{Wavelengths} &  \colhead{HD 209458} &
\colhead{$\epsilon$ Eri} & \colhead{HD 189733} & 
\colhead{GJ 436} & \colhead{GJ 876}\\
\colhead{Data Set} & \colhead{G0 V} & \colhead{K1 V} & \colhead{K1 V} &
\colhead{M3 V} & \colhead{M5.0 V}}
\startdata
\underline{$\Delta\lambda$=10--20 nm} & & & & &\\
\hspace{3mm}Model        & --1.046   & --1.025 & --1.065 & --0.583 & --0.560\\
\hspace{3mm}Sanz-Forcada & $<-2.625$ & --1.092 & --0.621 & --2.385 & --1.621\\
\hspace{3mm}EUVE\tablenotemark{a} &  & --1.063  &        &         &        \\
\underline{$\Delta\lambda$=20--30 nm} & & & & &\\
\hspace{3mm}Model        & --1.127  & --1.126 & --1.127 & --0.909 & --0.945\\
\hspace{3mm}Sanz-Forcada &$<-1.945$ & --0.822 & --0.661 & --1.985 & --1.411\\
\hspace{3mm}EUVE\tablenotemark{a} & & --0.909 &         &         &        \\
\underline{$\Delta\lambda$=30--40 nm} & & & & &\\
\hspace{3mm}Model        & --1.153  & --1.186 & --1.124 & --1.003 & --1.029\\
\hspace{3mm}Sanz-Forcada &$<-1.575$ & --0.972 & --0.681 & --1.435 & --1.051\\
\hspace{3mm}EUVE\tablenotemark{a} & & --1.046 &         &         &        \\
\underline{$\Delta\lambda$=40--70 nm} & & & & &\\
\hspace{3mm}Model        & --1.374 & --1.374 & --1.374 & --1.074 & --1.074\\
\hspace{3mm}Sanz-Forcada &$<-1.481$& --1.195 & --0.896 & --1.306 & --0.988\\
\underline{$\Delta\lambda$=70--91.2 nm} & & & & &\\
\hspace{3mm}Model        & --1.237 & --1.237 & --1.237 & --0.942 & --0.942\\
\hspace{3mm}Sanz-Forcada &$<-1.295$& --1.290 & --0.921 & --1.165 & --0.861\\
\underline{$\Delta\lambda$=10--91.2 nm} & & & & &\\
\hspace{3mm}Model        & --0.475 & --0.522 & --0.474 & --0.166 & --0.167\\
\hspace{3mm}Sanz-Forcada &$<-0.905$& --0.342 & --0.041 & --0.775 & --0.411\\
\underline{$\Delta\lambda$=91.2--117 nm} & & & & &\\
\hspace{3mm}Model        & --0.926 & --0.917 & --0.934 & --0.991 & --1.029\\
\hspace{3mm}X-exoplanets & --2.482 & --2.006 & --1.688 & --1.906 & --1.619\\
\hspace{3mm}FUSE+Ly series&        & --1.122 &         &         &        \\
\enddata
\tablenotetext{a}{{\em EUVE} fluxes of $\epsilon$~Eri are corrected for 
interstellar absorption using log[N(H~I)]=17.88 and \citet{Morrison1983}.}
\end{deluxetable}

\end{document}